\documentclass[nonacm,sigplan,screen]{acmart}
\usepackage{xspace}
\usepackage{listings}
\usepackage{xcolor}

\AtBeginDocument{%
  \providecommand\BibTeX{{%
    \normalfont B\kern-0.5em{\scshape i\kern-0.25em b}\kern-0.8em\TeX}}}

\setcopyright{none}

\begin{document}

\lstset{
    language=ml,
    basicstyle=\footnotesize\tt,      
    breaklines=true,
    commentstyle=\color{gray},
    morekeywords={mapping,enum,union, function,match},
    frame=single
}
\title{SIMD Everywhere Optimization from ARM NEON to RISC-V Vector Extensions}


\author{Ju-Hung Li}
\email{jhlee@pllab.cs.nthu.edu.tw}
\affiliation{%
  \institution{National Tsing Hua University}
  \country{Taiwan}
}

\author{Jhih-Kuan Lin}
\email{jklin@pllab.cs.nthu.edu.tw}
\affiliation{%
  \institution{National Tsing Hua University}
  \country{Taiwan}
}

\author{Yung-Cheng Su}
\email{ycsu@pllab.cs.nthu.edu.tw}
\affiliation{%
  \institution{National Tsing Hua University}
  \country{Taiwan}
}

\author{Chi-Wei Chu}
\email{cwchu@pllab.cs.nthu.edu.tw}
\affiliation{%
  \institution{National Tsing Hua University}
  \country{Taiwan}
}

\author{Lai-Tak Kuok}
\email{ltkuok@pllab.cs.nthu.edu.tw}
\affiliation{%
  \institution{National Tsing Hua University}
  \country{Taiwan}
}

\author{Hung-Ming Lai}
\email{hmlai@pllab.cs.nthu.edu.tw}
\affiliation{%
  \institution{National Tsing Hua University}
  \country{Taiwan}
}

\author{Chao-Lin Lee}
\email{clli@pllab.cs.nthu.edu.tw}
\affiliation{%
  \institution{National Tsing Hua University}
  \institution{PeakHills Group}
  \country{Taiwan}
}

\author{Jenq-Kuen Lee}
\email{jklee@cs.nthu.edu.tw}
\affiliation{%
  \institution{National Tsing Hua University}
  \country{Taiwan}
}


\begin{abstract}
Many libraries, such as  OpenCV, FFmpeg, XNNPACK, and Eigen, utilize Arm or x86 SIMD Intrinsics to optimize programs for performance.  With the emergence of RISC-V Vector Extensions (RVV), there is a need to migrate these performance legacy codes for RVV. Currently, the migration of NEON code to RVV code requires manual rewriting, which is a time-consuming and error-prone process. In this work, we use the open source tool,  "SIMD Everywhere" (SIMDe), to automate the migration. Our primary task is to enhance SIMDe to enable the conversion of ARM NEON Intrinsics types and functions to their corresponding RVV Intrinsics types and functions. For type conversion, we devise strategies to convert Neon Intrinsics types to RVV Intrinsics by considering the
vector length agnostic (vla) architectures. With function conversions, we analyze commonly used conversion methods in SIMDe and develop customized conversions for each function based on the results of RVV code generations. In our experiments with Google XNNPACK library, our enhanced SIMDe achieves speedup ranging from 1.51x to 5.13x compared to the original SIMDe, which does not utilize customized RVV implementations for the conversions.
\end{abstract}

\maketitle
\section{Introduction}

\begin{figure}[ht]   
    \centering  
    \includegraphics[width=8cm]{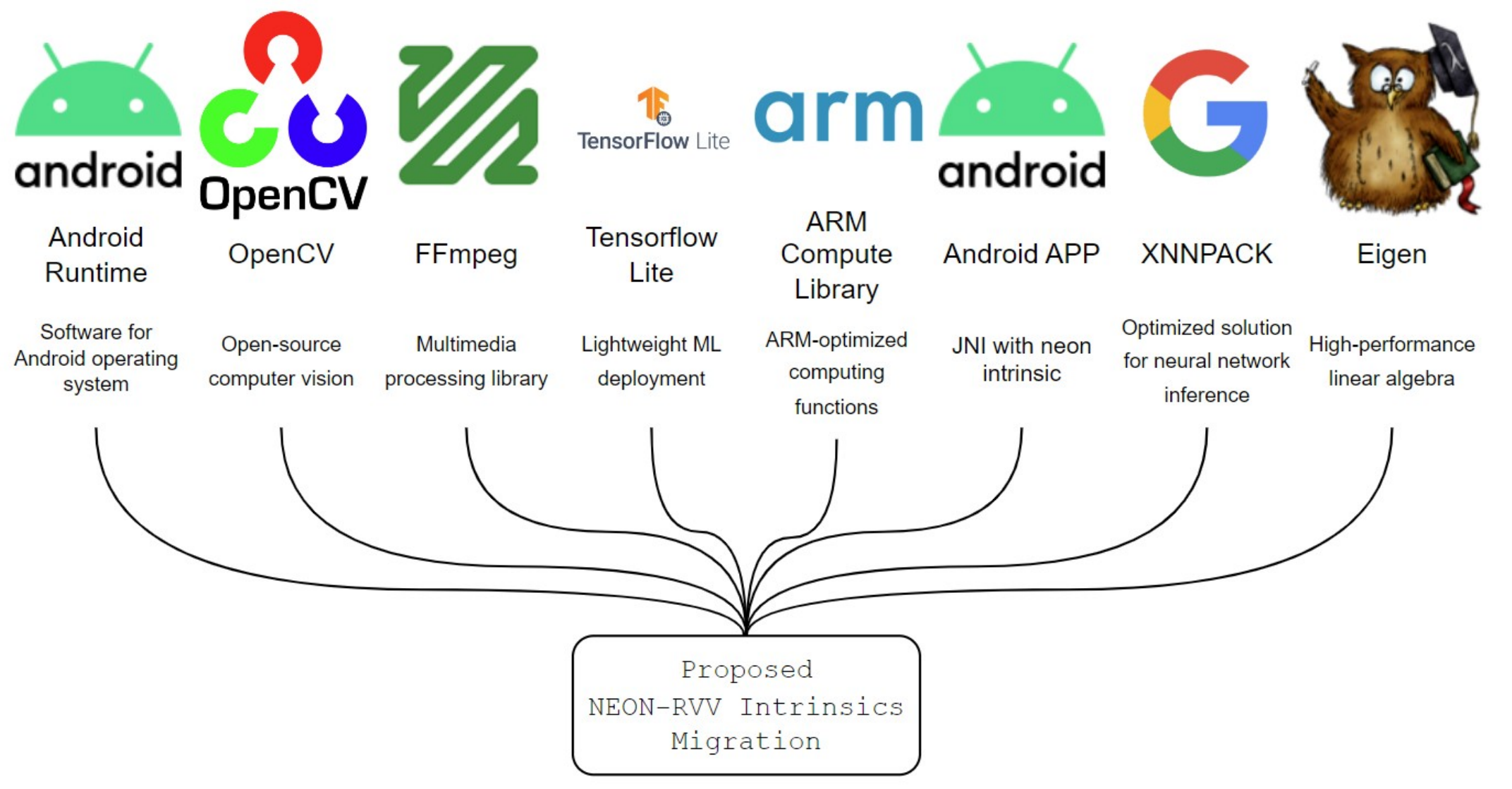}
    \caption{Key Applications Benefit from the Optimized Intrinsics Migration Flow for RVV}    
    \label{application}
\end{figure}

Many libraries, such as ComputeLibrary~\cite{compute}, OpenCV~\cite{opencv}, FFmpeg~\cite{ffmpeg}, XNNPACK~\cite{xnnpack}, and Eigen~\cite{eigen}, utilize Arm or x86 SIMD Intrinsics to optimize specific core algorithms and leverage the parallel processing capabilities of SIMD. With the emergence of RISC-V Vector Extensions (RVV)~\cite{rvv}, there is a need to migrate these libraries and legacy codes to take advantage of RVV instructions for improved performance on RISC-V platforms. Figure~\ref{application} illustrates key applications that can benefit from  the migration flow of Neon intrinsics to RVV intrinsics~\cite{rvv-intrinsics}. Our migration from ARM NEON Intrinsics to RISC-V Vector Extensions is expected to enhance several key applications. Android Runtime (ART) could witness enhanced system efficiency. Libraries like OpenCV and FFmpeg could experience faster processing times for computer vision tasks and multimedia data processing respectively. TensorFlow Lite could see improved execution speed, crucial for edge device deployment. Other Android applications could also see performance improvements. Machine learning libraries like XNNPACK could benefit in terms of on-device task performance, and the Eigen library could see improved calculation efficiency. In essence, this migration strategy is poised to drive significant enhancements across a range of applications in the Android ecosystem. Currently, the migration of NEON code to RVV code requires manual rewriting. Manual rewriting requires a good understanding of the architectural differences between the two instruction sets. It involves carefully modifying the instructions and data types in the code, which can be time-consuming, especially for larger codebases or when utilizing multiple Intrinsics. In this work, we attempt to automate the rewriting process with the open source tool, SIMD Everywhere (SIMDe). 

In this paper, we devise strategies to convert Neon Intrinsics types to RVV Intrinsics types. We also analyze commonly used conversion methods in SIMDe and develop customized conversions for each function based on the results of RVV code generation.  Neon Intrinsics types have lengths of 64 bits and 128 bits, while the type length of RVV Intrinsics is determined by the hardware implementation. This poses challenges in directly substituting Neon Intrinsics types with RVV Intrinsics types. Currently, SIMDe project does not yet have an implementation for converting instruction sets to the vector length agnostic (vla) architecture. Additionally, there are many Intrinsics in Neon that can not be directly replaced one-to-one with RVV Intrinsics. That makes it a challenge to effectively utilize RVV Intrinsics to achieve the functionality of Neon Intrinsics. In our work, we adopt the new
proposal from LLVM that D145088~\cite{fixed} proposes a fixed-size attribute for RISC-V Vector Extensions types , which allows declaring fixed-length RVV Intrinsics types given the length of a single register, making it easier to map NEON types to RVV types. Overall, we predominantly use customized RVV Intrinsics implementations for the conversions. The experiment shows our SIMDe achieved speedup ranging from 1.51x to 5.13x compared to the original SIMDe, which did not utilize customized RVV implementations for the conversions, when using XNNPACK as our benchmark. This work builds upon our prior efforts to enhance RISC-V software environments~\cite{yang2023auto, scan, lai2022efficient, hybrid, sail_paper, lin2021accelerate, subword, chen2020experiments}. Our previous work sets the stage for the next step in advancing the RISC-V ecosystem.

The remainder of the paper is organized as follows. In Section 2, we first introduce the Neon and RVV instruction sets. Next, in Section 3.1, we introduce the SIMD Everywhere design pattern for intrinsics function and type conversion. Next, we detail our strategies for leveraging SIMD Everywhere to migrate ARM NEON Intrinsics to RISC-V Vector Extensions Intrinsics in Sections 3.2 and 3.3. We explain how to use SIMD Everywhere for code porting in Section 3.4. Finally, in Section 4, we present the experimental results, comparing the native SIMDe with our RVV-enhanced SIMDe.

\section{Background}

\subsection{Neon}

Arm Neon is an single instruction multiple data (SIMD) architecture extension for the Arm Cortex-A and Arm Cortex-R series of processors with capabilities that vastly improve use cases on mobile devices, such as multimedia encoding/decoding, user interface, 2D/3D graphics, and gaming. Arm Neon has a total of 4344 Intrinsics. Table 1 categorizes the total number of Intrinsics by their Return base type.

\begin{table}
\begin{tabular}[ht]{|c|c|}
\hline
Return base type & Intrinsics counts\\
\hline
int & 1279\\
uint & 1448\\
float & 834\\
poly & 371\\
void & 331\\
bfloat & 81\\
\hline
\end{tabular}
\caption{Categorization of Neon Intrinsics with types}
\end{table}

\subsection{RISC-V Vector Extensions}

RISC-V Vector Extension (RVV) is an optional addition to the base RISC-V ISA, providing parallel computing capabilities. Unlike the RISC-V P extension~\cite{rvp}, which uses general-purpose registers (GPR) for packed-SIMD execution, RVV introduces a separate vector register file with 32 registers dedicated to SIMD operations. One notable feature of RVV is its flexibility in defining the vector length. Instead of being a fixed architectural constant, the vector length is determined by the implementation, allowing different microarchitectures to have varying vector lengths. This flexibility enables RVV programs to automatically scale across different implementations without the need for recompilation or rewriting.

\section{Migration Strategies}

\subsection{SIMD Everywhere}

SIMD Everywhere (SIMDe)~\cite{simde} is a header-only library designed to convert SIMD code across different architectures. It enables rapid transformation of SIMD libraries, enhancing the portability of SIMD code and reducing the time required for programmers to port SIMD libraries. SIMDe leverages the reuse of types and function conversions between various architectures, employing a general conversion approach. Regarding type conversions, SIMDe utilizes a union as a generic type. Within this union, besides the architecture-specific simd vector type variables, there is also a declaration of a type that is universally applicable across different architectures, typically an array or a variable with vector attributes. For example, Listing 1 is a universal int32x4 union used to convert NEON to other implementations. When the target ISA is SSE2, NEON, or WebAssembly, the union declares the corresponding simd vector types. Additionally, in all cases, a variable with vector attributes is also declared. In this work, we enhance SIMDe with the RVV target.

\begin{lstlisting}[caption=int32x4 union,captionpos=b,float]
typedef union {
  int32_t values __attribute__((__vector_size__(16)));
  #if defined(SIMDE_X86_SSE2_NATIVE)
    __m128i m128i;
  #endif
  #if defined(SIMDE_ARM_NEON_A32V7_NATIVE)
    int32x4_t neon;
  #endif
  #if defined(SIMDE_WASM_SIMD128_NATIVE)
    v128_t v128;
  #endif
} simde_int32x4_private;
\end{lstlisting}

As for function conversions, SIMDe employs specific ISA intrinsics for conversion and also utilizes compiler-specific vector extensions, built-in functions, and auto vectorization hints in a general conversion approach to enhance the portability of the SIMDe framework. For example, in Listing 2, the code converts Neon Intrinsics to other implementations. If the target ISA is Neon, AltiVec, SSE2, or WebAssembly, Neon Intrinsic vaddq\_s32 is transformed into the corresponding ISA implementation. If the target ISA is not one of the aforementioned options, the code utilizes variables with vector attributes for computations or auto vectorizes the scalar implementation. In our case for RVV, we can utilize LLVM backend for RVV so that the auto-vectorization flow can be obtained with a baseline solution. We further enhance the flow with RVV intrinsics in the transformation.

\begin{lstlisting}[caption=Neon intrinsics vaddq\_s32 conversion,captionpos=b]
SIMDE_FUNCTION_ATTRIBUTES
simde_int32x4_t
simde_vaddq_s32(simde_int32x4_t a, simde_int32x4_t b) {
  #if defined(SIMDE_ARM_NEON_A32V7_NATIVE)
    return vaddq_s32(a, b);
  #elif defined(SIMDE_POWER_ALTIVEC_P6_NATIVE)
    return vec_add(a, b);
  #else
    simde_int32x4_private
      r_,
      a_ = simde_int32x4_to_private(a),
      b_ = simde_int32x4_to_private(b);

    #if defined(SIMDE_X86_SSE2_NATIVE)
      r_.m128i = _mm_add_epi32(a_.m128i, b_.m128i);
    #elif defined(SIMDE_WASM_SIMD128_NATIVE)
      r_.v128 = wasm_i32x4_add(a_.v128, b_.v128);
    #elif defined(SIMDE_VECTOR_SUBSCRIPT_OPS)
      r_.values = a_.values + b_.values;
    #else
      clang loop vectorize(enable)
      for (size_t i = 0 ; i < (sizeof(r_.values) / sizeof(r_.values[0])) ; i++) {
        r_.values[i] = a_.values[i] + b_.values[i];
      }
    #endif

    return simde_int32x4_from_private(r_);
  #endif
}

\end{lstlisting}

\subsection{Migration Strategies with Type Conversion}

Neon Intrinsics types have lengths of 64 bits and 128 bits, while the type length (vlen) of RVV Intrinsics is determined by the hardware implementation. This makes it difficult to directly substitute Neon Intrinsics types with RVV Intrinsics types. Additionally, because RVV vlen is known at runtime, RVV Intrinsics types are sizeless types. Sizeless types have greater limitations, such as not being able to be declared in global variables, structs, or unions. This poses a challenge when replacing Neon Intrinsics types within specific areas. To address the inconvenience of sizeless types, LLVM recently introduced a new attribute for RVV Intrinsics types. With this attribute, RVV Intrinsics types can be treated as fixed-size vectors when the architecture's vlen is known.

To perform type conversion, we modify simde/arm/neon/types.h so that it includes corresponding RVV Intrinsics types. Since Neon Intrinsics types have lengths of 64 bits and 128 bits, we consider that for effective substitution, RVV vlen should be at least 64 bits for replacing Neon 64-bit types, and at least 128 bits for replacing Neon 128-bit types. This allows for substitution without relying on loops for operations. Additionally, in RVV, the number of processed elements is determined by setting vector length register vl. RVV vlen only restricts the maximum number of processed elements and does not solely determine it. Therefore, as long as the RVV vlen is greater than the vector length of Neon, type substitution can be performed. Listing 3 shows the code that adds RVV types to Neon generic int32x4 type. The variable "\_\_riscv\_v\_fixed\_vlen" determines the length of RVV vector, which is currently determined by the compiler flag.

\begin{lstlisting}[caption=int32x4 union with RVV type,captionpos=b]
typedef vint32m1_t fixed_vint32m1_t __attribute__((riscv_rvv_vector_bits(__riscv_v_fixed_vlen)));

typedef union {
    ...
  #if defined(SIMDE_RISCV_V_NATIVE) && SIMDE_NATURAL_VECTOR_SIZE >= 128
    fixed_vint32m1_t sv128;
  #endif
    ...
} simde_int32x4_private;
\end{lstlisting}

Since the size of the union depends on the size of the largest variable, when the size of the Neon type is smaller than the size of the RVV type, the size of the union increases. Currently, in the conversion implementation of store intrinsics in SIMDe, memcpy is used to copy a number of bytes equal to the size of the union from the memory location of the union to the destination memory address. This can lead to errors in SIMDe during partial conversions. Therefore, regardless of the quality of SIMDe's original implementation, in this scenario, we use customized RVV Intrinsics implementation to correctly store the desired number of elements in memory, as shown in the code in Listing 4.

\begin{lstlisting}[caption=Neon intrinsics vst1q\_s32 conversion,captionpos=b]
void
simde_vst1q_s32(int32_t ptr[HEDLEY_ARRAY_PARAM(4)], simde_int32x4_t val) {
    ...
    simde_int32x4_private val_ = simde_int32x4_to_private(val);
    #if defined(SIMDE_RISCV_V_NATIVE) && (SIMDE_NATURAL_VECTOR_SIZE >= 128)
      __riscv_vse32_v_i32m1(ptr , val_.sv128 , 4); // Ensure that we save the correct number of elements into memory.
    #else
      simde_memcpy(ptr, &val_, sizeof(val_));
    #endif
  #endif
}
\end{lstlisting}

If there is no corresponding RVV type for a Neon type, the RVV type variable is not declared in the union. Possible scenarios include: 

\begin{enumerate}
    \item If vlen is less than 64 bits, substitution is not straightforward  for Neon 64-bit types.
    \item If vlen is less than 128 bits, substitution is not straightforward for Neon 128-bit types.
    \item Without the Zvfh extension, f16 vectors can not be replaced straightforwardly in RVV. 
\end{enumerate}

In cases where substitution is not possible, the variables with vector attribute in the union can still be used for intrinsics conversion. Table 2 is the mapping table for RVV and Neon types, assuming that the Zvfh extension is enabled. LLVM D145088 proposes a fixed-length attribute for RVV intrinsics type with LMUL=1. Therefore, we currently use RVV intrinsics type with LMUL=1 for the conversion.

\begin{table}
\begin{tabular}[!htpb]{|l|l|l|l|}
\hline
Neon & vlen<64 & 64<=vlen<128 & vlen>=128 \\
\hline
int8x8\_t   & x & vint8m1\_t & vint8m1\_t\\
int16x4\_t   & x & vint16m1\_t & vint16m1\_t\\
int32x2\_t   & x & vint32m1\_t & vint32m1\_t\\
int64x1\_t   & x & vint64m1\_t & vint64m1\_t\\
uint8x8\_t   & x & vuint8m1\_t & vuint8m1\_t\\
uint16x4\_t   & x & vuint16m1\_t & vuint16m1\_t\\
uint32x2\_t   & x & vuint32m1\_t & vuint32m1\_t\\
uint64x1\_t   & x & vuint64m1\_t & vuint64m1\_t\\
float16x4\_t   & x & vfloat16m1\_t & vfloat16m1\_t\\
float32x2\_t   & x & vfloat32m1\_t & vfloat32m1\_t\\
float64x1\_t   & x & vfloat64m1\_t & vfloat64m1\_t\\
int8x16\_t   & x & x & vint8m1\_t\\
int16x8\_t   & x & x & vint8m1\_t\\
int32x4\_t   & x & x & vint8m1\_t\\
int64x2\_t   & x & x & vint8m1\_t\\
uint8x16\_t   & x & x & vint8m1\_t\\
uint16x8\_t   & x & x & vint8m1\_t\\
uint32x4\_t   & x & x & vint8m1\_t\\
uint64x2\_t   & x & x & vint8m1\_t\\
float16x8\_t   & x & x & vfloat16m1\_t\\
float32x4\_t   & x & x & vfloat32m1\_t\\
float64x2\_t   & x & x & vfloat64m1\_t\\
\hline
\end{tabular}
\caption{Mapping table for Neon types and RVV types with fixed-size attribute}
\end{table}

\subsection{Migration Strategies with Function Conversion}

Here are five commonly used conversion methods in the SIMDe framework.

\begin{enumerate}
    \item Utilizing ISA-specific Intrinsics functions
    \item Utilizing vector built-in functions.
    \item Performing vector operations utilizing variables with vector attributes.
    \item Vectorizing scalar implementations through the compiler's auto-vectorization pass.
    \item Combine other functions.
\end{enumerate}

Currently, SIMDe has implementations for converting Neon intrinsics to generic architecture code. However, since there is no specific implementation for converting Neon Intrinsics to RVV Intrinsics, it can only utilize vector attributes or use compiler auto-vectorization pass implementations. 

Unfortunately, these methods are unable to generate the optimal RVV code in many cases. Overall, in our design, we present customized RVV Intrinsics implementations for the conversions and have implemented conversions for a total of 1520 Intrinsics. RVV has many Intrinsics that have the same functionality as Neon Intrinsics and can be directly substituted one to one. For some Intrinsics, by combining a few RVV Intrinsics, we can achieve the same functionality as the corresponding Neon Intrinsics. For example, Listing 5 provides the code for converting Neon's vget\_high\_s32 using a customized RVV Intrinsics implementation. Neon "get\_high" Intrinsics is used to extract the upper N/2 elements from a vector of width N and generate a new vector of width N/2. We replaced it with RVV "slidedown" Intrinsics, which shifts the vector elements down by a specified number of positions. 

\begin{lstlisting}[caption=Neon intrinsics vget\_high\_s32 conversion,captionpos=b,float]
simde_int32x2_t
simde_vget_high_s32(simde_int32x4_t a) {
    ...
    simde_int32x2_private r_;
    simde_int32x4_private a_ = simde_int32x4_to_private(a);
    #if defined(SIMDE_RISCV_V_NATIVE) 
    && (SIMDE_NATURAL_VECTOR_SIZE >= 128)
      r_.sv64 = __riscv_vslidedown_vx_i32m1(a_.sv128 , 2 , 4);
    ...
    return simde_int32x2_from_private(r_);
  #endif
}
\end{lstlisting}

Listing 6 presents another example for converting Neon's vceqq\_s32 using a customized RVV Intrinsics implementation. Neon "ceq" Intrinsics compare two vectors, and if the corresponding elements are equal, it sets all the bits of the corresponding elements in the result vector to 1; otherwise, it sets them to 0. We achieve the same functionality by combining different RVV instructions. In this process, the vmv instruction is used to generate a vector, vs\_0, with all elements set to 0. Then, the vmseq instruction compares the corresponding elements of the two vectors and generates a mask vector. Finally, the vmerge function combines vs\_0 and -1 based on the mask vector, resulting in the final output vector. 

\begin{lstlisting}[caption=Neon intrinsics vceqq\_s32 conversion,captionpos=b]
vuint8m1_t mask;
simde_vceqq_s32(simde_int32x4_t a, simde_int32x4_t b) {
    ...
    simde_uint32x4_private r_;
    simde_int32x4_private
      a_ = simde_int32x4_to_private(a),
      b_ = simde_int32x4_to_private(b);
    #if defined(SIMDE_RISCV_V_NATIVE)
     && (SIMDE_NATURAL_VECTOR_SIZE >= 128)
      vuint32m1_t vs_0 = __riscv_vmv_v_x_u32m1(UINT32_C(0), 4);
      vbool32_t mask = __riscv_vmseq_vv_i32m1_b32(a_.sv128, b_.sv128, 4);
      r_.sv128 = __riscv_vmerge_vxm_u32m1(vs_0, -1, mask, 4);
    ...
    return simde_uint32x4_from_private(r_);
  #endif
}
\end{lstlisting}

We now give an example that Neon Intrinsics may require more complex conversions. This example is for Neon 'rbit' Intrinsics, which reverses the bit order of each element in a vector. To achieve the same functionality using RVV, we refer to Edwin Freed's article 'Binary Magic Numbers' from Dr. Dobb's Journal 1983~\cite{bitreverse} for the bit reverse solution. Listing 7 provides code implementing an algorithm that reverses the bit order in an unsigned integer 'v'. It accomplishes the result by swapping odd and even bits, consecutive pairs of bits, nibbles (groups of 4 bits), bytes (groups of 8 bits), and 2-byte long pairs through a series of bitwise operations and shifts. We implement a SIMD version of this algorithm using RVV bitwise operation intrinsics.

\begin{lstlisting}[caption=Bit reverse solution,captionpos=b]
v = ((v >> 1) & 0x55555555) | ((v & 0x55555555) << 1);
v = ((v >> 2) & 0x33333333) | ((v & 0x33333333) << 2);
v = ((v >> 4) & 0x0F0F0F0F) | ((v & 0x0F0F0F0F) << 4);
v = ((v >> 8) & 0x00FF00FF) | ((v & 0x00FF00FF) << 8);
v = ( v >> 16             ) | ( v               << 16);
\end{lstlisting}

While we primarily use customized RVV implementations, we also retain the use of vector attributes in certain cases. These cases include situations where Neon Intrinsics types in the parameters can not be replaced with RVV types, as well as Intrinsics that are specifically designed for simple vector arithmetic or shift operations. Using vector attributes for such Intrinsics often leads to optimal RVV code generation. This ensures that we can produce optimal RVV code and maximize the reuse of conversions in SIMDe. Listing 8 provides the code for converting Neon's simde\_vaddq\_s32 using variables with vector attributes.

\begin{lstlisting}[caption=Neon intrinsics vaddq\_s32 conversion,captionpos=b]
SIMDE_FUNCTION_ATTRIBUTES
simde_int32x4_t
simde_vaddq_s32(simde_int32x4_t a, simde_int32x4_t b) {
    ...
    simde_int32x4_private
      r_,
      a_ = simde_int32x4_to_private(a),
      b_ = simde_int32x4_to_private(b);
    ...
    #elif defined(SIMDE_VECTOR_SUBSCRIPT_OPS)
      r_.values = a_.values + b_.values;
    return simde_int32x4_from_private(r_);
  #endif
}
\end{lstlisting}

\subsection{SIMDe Usage}

The usage of SIMDe is straightforward. Simply includes the SIMDe header file in the code that needs to be converted. In Listing 9, the code uses our RVV-enhanced SIMDe header file to convert the vector addition code implemented with Neon Intrinsics, we include neon.h header file in line 3. After compilation, it produces the RVV code shown in Listing 10. In lines 7 and 10, the Neon vld1q\_s32 instruction is converted to the RVV vle32 instruction. In line 11, the Neon vaddq\_s32 instruction is converted to the RVV vadd instruction. Finally, in line 12, the Neon vst1q\_s32 instruction is converted to the RVV vse32 instruction.

\begin{lstlisting}[caption=Neon vector addition ,captionpos=b , numbers=left,float]
#include <stdio.h>
#define SIMDE_ENABLE_NATIVE_ALIASES
#include "simde/arm/neon.h" // header file for conversion Neon to RVV

int A[] = { 0 , 1 , 2 , 3 };
int B[] = { 4 , 5 , 6 , 7 };
int main(void){
        int32x4_t va , vb , vc;
        va = vld1q_s32(A);
        vb = vld1q_s32(B);
        va = vaddq_s32(va , vb);
        vst1q_s32(A , va);
        printf("%d" , A[0]);
        return 0;
}
\end{lstlisting}

\begin{lstlisting}[caption=RVV vector addition ,captionpos=b , numbers=left,float]
<main>:
sp,sp,-16
sd      ra,8(sp)
lui     a0,0x1e
add     a1,a0,1800 # 1e708 <A>
vsetivli        zero,4,e32,m1,ta,ma
vle32.v v8,(a1) // va = vld1q_s32(A);
lui     a2,0x1e
add     a2,a2,1816 # 1e718 <B>
vle32.v v9,(a2) // vb = vld1q_s32(B);
vadd.vv v8,v8,v9 // va = vaddq_s32(va , vb);
vse32.v v8,(a1) // vst1q_s32(A , va);
lw      a1,1800(a0)
lui     a0,0x1c
add     a0,a0,2000 # 1c7d0 <__clzdi2+0x42>
jal     1034c <printf>
li      a0,0
ld      ra,8(sp)
add     sp,sp,16
ret
\end{lstlisting}

\section{Experiments}

\subsection{Validation Workflow}

SIMDe includes unit tests for converting Neon code to other ISAs. These unit tests validate each instruction using multiple test cases to ensure the conversion functions correctly under different scenarios. We reused the unit tests within SIMDe and validated them using Spike simulator.

\subsection{Benchmark Experiments}

We use XNNPACK as our benchmark. XNNPACK is an open-source software library developed by the Google team, aims to optimize neural network computations across different hardware platforms. Within XNNPACK, there are various neural network computation functions implemented using NEON Intrinsics.

Experiments were conducted to compare native SIMDe with our implementation. We specifically chose 10 commonly used neural network computation functions that are implemented using NEON Intrinsics in XNNPACK. Below are brief descriptions of the 10 functions. Gemm is a high-performance function for general matrix multiplication. Conv-hwc is a convolution function specifically for input data arranged in the Height-Width-Channel format. Dwconv is optimized for depthwise separable convolution with reduced parameters and computation. Maxpool extracts the maximum value from each region during pooling. Argmaxpool performs maxpooling while returning the index of the maximum value. Vrelu applies the ReLU activation function element-wise. Vsqrt calculates the square root of each element in the input vector. Vtanh applies the hyperbolic tangent activation function element-wise. Vsigmoid applies the sigmoid activation function element-wise. Lastly, Ibilinear is a high-performance function specialized in performing bilinear interpolation.

The SIMDe header file was included in these functions, and the code was compiled using Clang compiler with O3 optimization level. During the preprocessing stage, the functions accelerated with NEON Intrinsics were transformed into RVV Intrinsics. Following compilation, we executed the executable file using Spike simulator to verify correctness and calculate the instruction counts. Since Spike is a functional model rather than a cycle-accurate simulator, we employed dynamic instruction count as the performance metric instead. In this experiment, we used Spike 1.1.1 and Clang 17.0.0 ( commit hash  5326c9e480d70e16c2504cb5143524aff3ee2605 ). Figure 2 illustrates our experimental results, indicating that SIMDe achieved speedup ranging from 1.51x to 5.13x compared to the original SIMDe, which did not utilize customized RVV implementations for the conversions. The original flow goes with clang vector attributes for computations or auto vectorization of the scalar
implementation. It then utilizes LLVM
backend for RVV so that the clang vector attribute and the auto-vectorization flow can be obtained as a baseline solution.

\begin{figure}[ht]   
    \centering  
    \includegraphics[width=9cm]{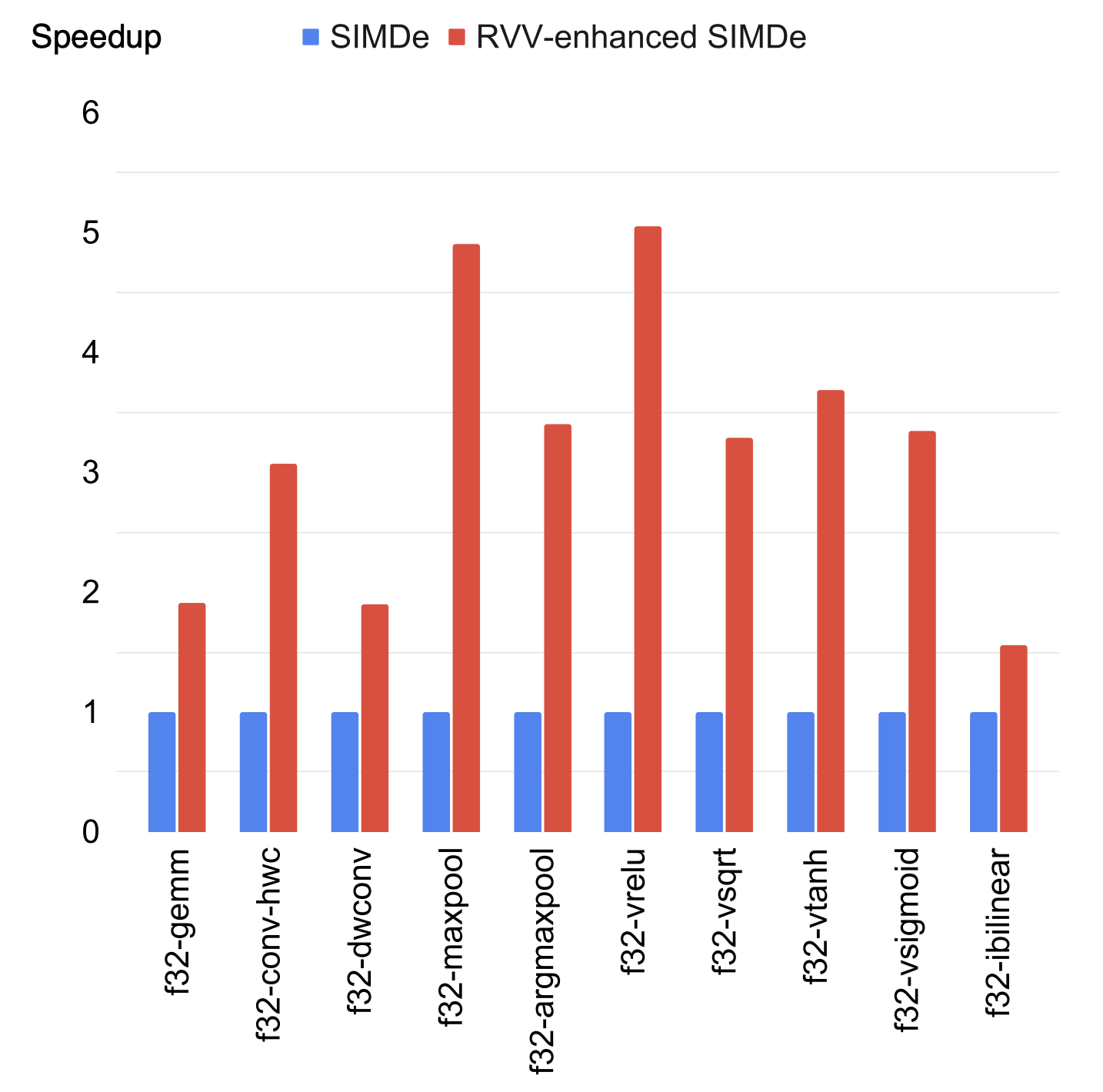}
    \caption{RVV-enhanced SIMDe Performance Comparison} 
\end{figure}

\section{Conclusion}

In this paper, we use the open source tool,  "SIMD Everywhere" (SIMDe), to automate the migration from Neon code to RVV code. Our primary task is to enhance SIMDe to enable the conversion of ARM NEON Intrinsics types and functions to their corresponding RVV Intrinsics types and functions. In our experiments with Google XNNPACK library, our RVV-enhanced SIMDe achieves speedup ranging from 1.51x to 5.13x compared to the original SIMDe, which does not utilize customized RVV implementations for the conversions.


\bibliographystyle{ACM-Reference-Format}
\bibliography{sample-base}

\end{document}